\documentclass[12pt,a4paper]{article}

\usepackage[british]{babel}

\usepackage[a4paper,top=2cm,bottom=2cm,left=2.5cm,right=2.5cm,marginparwidth=1.75cm]{geometry}

\usepackage[style=apa, backend=biber]{biblatex} % APA 7th edition style citations using biblatex
\DeclareLanguageMapping{british}{british-apa} % Set language mapping
\DeclareFieldFormat[article]{volume}{\apanum{#1}} % Format volume number

\usepackage{amsmath}
\usepackage{graphicx}
\usepackage[colorlinks=true, allcolors=blue]{hyperref}
\usepackage{hyperref}
\usepackage[title]{appendix}
\usepackage{mathrsfs}
\usepackage{amsfonts}
\usepackage{booktabs} % For \toprule, \midrule, \botrule
\usepackage{caption}  % For \caption
\usepackage{threeparttable} % For table footnotes
\usepackage{algorithm}
\usepackage{algorithmicx}
\usepackage{algpseudocode}
\usepackage{listings}
\usepackage{enumitem}
\usepackage{chngcntr}
\usepackage{booktabs}
\usepackage{lipsum}
\usepackage{subcaption}
\usepackage{authblk}
\usepackage[T1]{fontenc}    % Font encoding
\usepackage{csquotes}       % Include csquotes
\usepackage{diagbox}

\usepackage{setspace}
\usepackage{titlesec}
\titleformat{\section} % Redefine section numbering format
  {\normalfont\Large\bfseries}{\thesection.}{1em}{}

\usepackage{float}   % for customizing caption position
\usepackage{caption} % for customizing caption format


\title{American Call Options Pricing With Modular Neural Networks}

\author{Ananya Unnikrishnan}

\date{September 29, 2024} 

\begin{document}

\maketitle

\begin{abstract}
An accurate valuation of American call options is critical in most financial decision making environments. However, traditional models like the Barone-Adesi Whaley (B-AW) and Binomial Option Pricing (BOP) methods fall short in handling the complexities of early exercise and market dynamics present in American options. This paper proposes a Modular Neural Network (MNN) model which aims to capture the key aspects of American options pricing. By dividing the prediction process into specialized modules, the MNN effectively models the non-linear interactions that drive American call options pricing. Experimental results indicate that the MNN model outperform both traditional models as well as a simpler Feed-forward Neural Network (FNN) across multiple stocks (AAPL, NVDA, QQQ), with significantly lower RMSE and nRMSE (by mean). These findings highlight the potential of MNNs as a powerful tool to improve the accuracy of predicting option prices.  
\end{abstract}

%--- Section ---%
\section{Introduction}
\label{sec:introduction}
In the dynamic landscape of financial markets, the ability to predict the price of call options has become crucial to effectively manage and reduce the risks associated with investment portfolios \parencite{dejanovski_role_2014}. Options allow traders and investors to hedge against downside risk as well as speculate on price movements. A call option gives the holder the right, but not the obligation, to buy an underlying asset at a specified price within a certain period, while a put option grants the right to sell the asset under similar conditions. American options in particular stand out as they can be exercised at any point before expiration, which makes them more flexible but also adds complexity to their valuation in comparison to European options, which can only be exercised at maturity \parencite{ekstrom_properties_2004}. While this flexibility is advantageous to investors, it however complicates the pricing of these options especially under dynamic market conditions. Traditional pricing models struggle to account for the non-linear relationships that exist in American options, even if they offer fundamental insights \parencite{tudor_comparison_2022}. 

Neural networks, which have transformed data-driven analysis in areas such as image recognition and natural language processing, offer a promising alternative to traditional pricing models for option pricing \parencite{alishahi_analyzing_2019}. With their ability to learn complex, non-linear relationships from huge datasets, neural networks can capture the intricacies of American options pricing, especially with their added layer of complexity from early exercise decisions. By letting the model learn to price from past records as well as present market information, they provide a more dynamic approach to pricing in comparison to traditional pricing models \parencite{dase_application_2010}. The purpose of this study is to explore the effectiveness of Modular Neural Networks (MNNs) in predicting American call options prices. By comparing the MNN model performance to traditional models such as the Barone-Adesi and Whaley (B-AW) and Binomial Option Pricing (BOP) methods as well as a simpler Feed-forward Neural Network (FNN) as a benchmark, this study aims to contribute to the growing body of research applying neural networks to financial modeling.

%--- Section ---%
\section{Literature Review}
\label{sec:literature_review}
The Barone-Adesi and Whaley (B-AW) method offers a analytical solution that allows for the computation of the price of American options while considering their early exercise aspect. The B-AW model was a significant advancement over the Black-Scholes (B-S) model, which is limited to pricing European options and lacks the capacity to handle the early exercise premium associated with American options \parencite{barone-adesi_efficient_1987}. However, despite its superior ability to predict American options in comparison to earlier models, the B-AW model is limited by several key assumptions, such as constant volatility which may not hold true in dynamic market environments \parencite{yang_whaley_2022}. 

The Binomial Option Pricing (BOP) model, first introduced by \textcite{cox_theory_1979}, offers a numerical method for determining the price of American and European options. The BOP model is based on the principle of constructing a binomial tree that corresponds to possible paths of the underlying asset’s price over time. This aspect is crucial for American options as it facilitates the early exercising of options at each node of the tree. While the BOP model is easy to implement, it can be quite hard to adapt to more complex situations like accounting for the value of managerial flexibility \parencite{fadugba_performance_2014}.

Neural networks perform exceptionally well in tasks that are challenging for traditional models due to their ability to draw insights from large amounts of data and accommodating diverse inputs. The use of neural networks in the pricing of options was first demonstrated for European options through the work of \textcite{hutchinson_nonparametric_1994}. They employed non-parametric techniques to prove that a neural network can learn to price European options like the S\&P 500 without any of the classical modeling constraints. Building on this ground, a number of studies have focused on bases of neural networks for modeling of European options. For instance, \textcite{amilon_neural_2003} further developed the research of Hutchinson et al. examining the neural networks application in Swedish index call options pricing and proving better performance in modeling complex relationships like bid-ask spreads than their Black-Scholes counterparts. Although Amilon's study focused on European options, it highlighted neural networks' adaptability in modeling the relationships between option prices and market variables. While the application of neural networks to American options has been relatively limited compared to European options, there have been promising studies in recent years. For example, \textcite{gaspar_neural_2020} used neural networks to price American put options and managed to show that they can produce even lower RMSE than the Least-Squares Monte Carlo (LSM) method across different U.S. companies.

While significant research has been done on neural networks for European options, such as \textcite{gradojevic_option_2009} work on using Modular Neural Networks (MNNs) for European option pricing, the literature on neural network models for American options remains comparatively sparse. This paper thus seeks to bridge this gap by applying MNNs to the problem of American call option pricing.

%--- Section ---%
\section{Methodology}
\label{sec:methodology}

\subsection{Feed-forward Neural Network Architecture}
\begin{figure}[ht]
 \centering
 \makebox[\textwidth][c]{\includegraphics[width=0.5\textwidth]{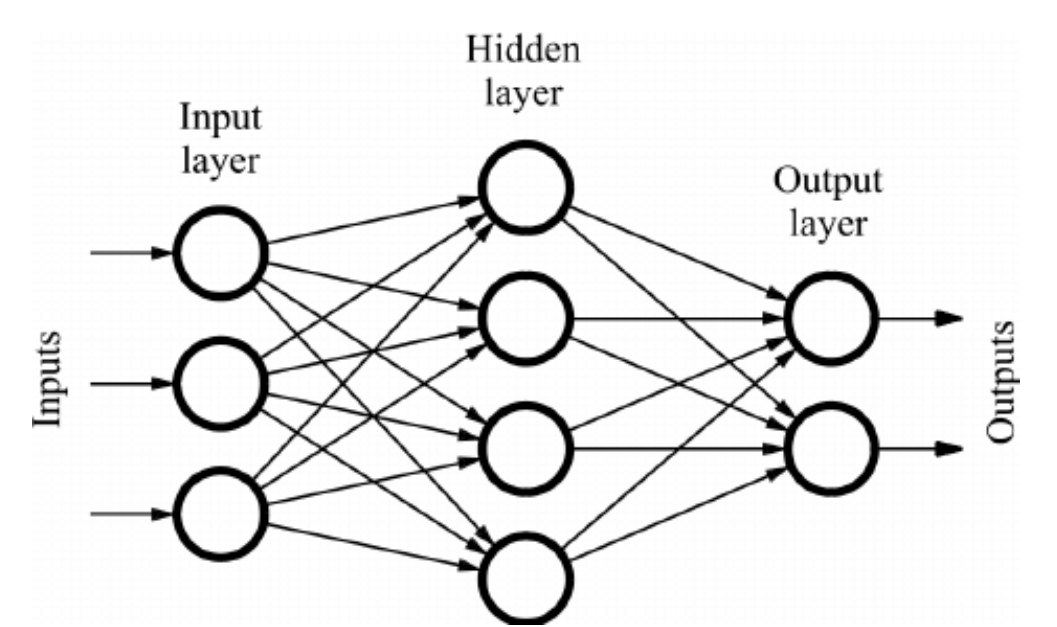}}
 \caption{FNN Architecture}
 \label{fig-1}
\end{figure}

Feed-forward Neural Networks (FNN) are one of the simplest architectures in artificial neural networks (ANN). It has three major layers: input layer, hidden (one or more) layers, and output layer. The neurons in the same layer are connected to the neurons in the next layer through weighted connections. This structure is known as ‘feed-forward’ since information moves in only one direction, from the input layer to the output layer, without any feedback loops as illustrated in Figure \ref{fig-1}. Every neuron processes the input it receives through an activation function which adds non-linearity to the network, making it capable of understanding complex patterns in data \parencite{laudani_training_2015}.

\subsection{Modular Neural Network Architecture}
\begin{figure}[ht]
 \centering
 \makebox[\textwidth][c]{\includegraphics[width=0.5\textwidth]{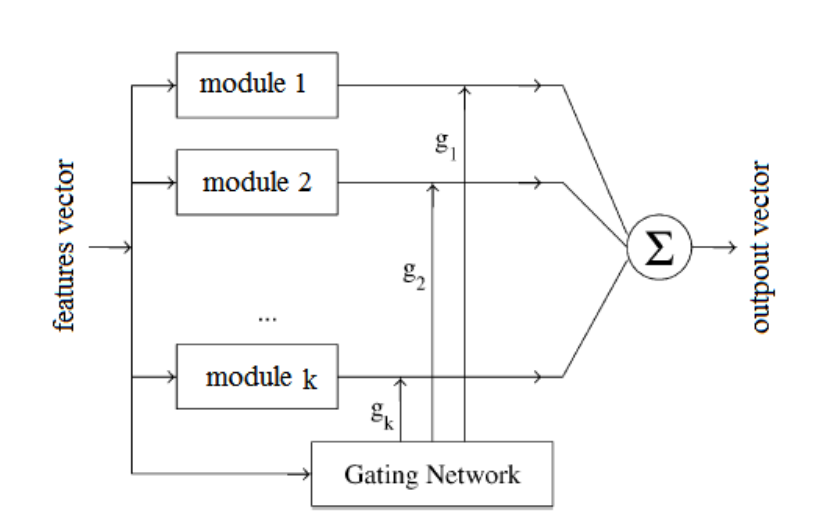}}
 \caption{MNN Architecture}
 \label{fig-2}
\end{figure}

Modular neural networks (MNNs) are a unique type of artificial neural network (ANN) which aim to solve complicated problems by dividing them into small independent modules. Each module takes care of a particular group of features or a sub-task which makes the overall network efficient in dealing with diverse and complex problems \parencite{auda_modular_1999}. By dividing a larger problem into smaller, manageable sub-problems as shown in Figure \ref{fig-2}, MNNs can focus on tackling specific aspects of the task independently.

The design of MNNs differs from classic feed-forward neural networks (FNNs) in one important aspect: while FNNs treat all inputs as one entire network, MNNs incorporate several standalone networks which use only a part of the inputs. Since MNNs have the structural advantage in handling specialization and complexity, they are especially well-suited for predicting options \parencite{gradojevic_option_2009}. Moreover, MNNs have the advantage of parallelization during training, as each module can be optimized independently. This not only reduces the time required to train large models but also ensures that each module achieves a high level of accuracy in its specialized task, contributing to a more accurate overall prediction.

\subsection{Proposed MNN Model Architecture}
In this study, we propose a Modular Neural Network (MNN) model specifically designed to predict American call option prices. Due to the added complexity of American options, specifically from the feature of the early exercise, it calls for a method that is capable of capturing the complex and diverse aspects of options pricing. The MNN architecture divides the prediction process into six specialized modules, each responsible for a particular set of features, which represent the different factors that affect an option’s price. By assigning specialized tasks to separate modules, MNNs are better equipped to capture the intricate relationships present in American call options.

We have introduced six modules that processes a specific subset of features that relate to distinct components of options pricing. They are as follows:
\begin{enumerate}
\item Intrinsic Value and Moneyness Module
\item Time Value and Volatility Module
\item Early Exercise and Dividend Module
\item Liquidity and Market Conditions Module
\item Macro-Economic and Sentiment Module
\item Option Greeks and Interaction Module
\end{enumerate}

The modularization of this architecture offer multiple benefits over traditional pricing models as well as standard neural networks. Firstly, each module specializes in a specific aspect of options pricing, allowing the network to capture complex, non-linear interactions between factors. Furthermore, by structuring the network in this way, the MNN not only improves prediction accuracy but also enhances the interpretability of the model. Each module corresponds to a well-understood financial concept, making it easier to understand how different factors influence the model’s predictions. Lastly, the MNN architecture is highly adaptable, allowing for hyper-parameter tuning in each module to optimize performance based on market conditions or specific datasets.

\subsubsection{Intrinsic Value and Moneyness Module}
The Intrinsic Value and Moneyness Module is designed to capture the effects of the underlying asset’s price and the option’s strike price on the option’s intrinsic value and moneyness. This is important because it helps in understanding the option's immediate value and its potential for early exercise. The features used in this module include:
\begin{itemize}
\item Underlying last price: This is the current market price of the underlying asset, a critical input in determining the option’s intrinsic value.
\item Strike price: The price at which the option holder can buy (call) or sell (put) the underlying asset, used to calculate the difference between the current asset price and the strike price.
\item Intrinsic call value: This represents the value the option would have if exercised immediately, calculated as the difference between the underlying asset price and the strike price for call options.
\item Moneyness: This feature captures the relative position of the underlying price to the strike price and is a key factor in early exercise decisions.
\item Log of moneyness: The logarithmic transformation of moneyness helps stabilize the variance and enables the model to better capture nonlinear relationships between asset prices and the option’s moneyness.
\item Moneyness multiplied by strike distance: This feature captures the combined effect of how far the option is in or out of the money relative to its strike distance, enhancing the module’s ability to predict option price behavior under varying market conditions.
\end{itemize}

\subsubsection{Time Value and Volatility Module}
The Time Value and Volatility Module is designed to capture how the time value diminishes and how volatility fluctuations impact the option price, both of which are essential for American options where timing is crucial. The features used in this module include:
\begin{itemize}
\item Days to expiration (DTE): The number of days until the option expires. This is crucial as time decay (theta) accelerates as expiration approaches.
\item Implied volatility: The market’s expectation of the future volatility of the underlying asset, which significantly impacts the premium of the option.
\item Historical volatility: A measure of the actual past volatility of the underlying asset, which helps in gauging whether the current implied volatility is in line with past behavior.
\item Realized volatility over 20 days: Captures the asset’s actual volatility over the recent past, providing a short-term view of price fluctuations.
\item Option theta: Theta measures the rate at which the option's value erodes as it approaches expiration, reflecting the time decay component.
\item Option gamma: Gamma measures the rate of change in the option’s delta, providing insight into how sensitive the option is to movements in the underlying asset.
\item Log of days to expiration: Taking the logarithm of DTE helps model the non-linear impact of time on the option price.
\item DTE multiplied by implied volatility: This interaction term helps capture the combined effect of time and volatility, which are closely intertwined in determining option value.
\item Gamma divided by DTE: This ratio gives insights into how option sensitivity changes with time, particularly relevant as options approach expiration.
\end{itemize}

\subsubsection{Early Exercise and Dividend Module}
The Early Exercise and Dividend Module is designed to capture the unique dynamics surrounding early exercise for dividend-paying stocks as holders may exercise early to capture dividend. The features used in this module include:
\begin{itemize}
\item Dividend yield: The annual dividend expressed as a percentage of the stock price, which incentivizes early exercise for call options.
\item Days to ex-dividend date: The number of days until the ex-dividend date, after which the stock trades without the value of the next dividend, prompting early exercise before this date.
\item Interest rate: The risk-free rate of return, which influences the cost of holding an option versus exercising it early.
\item Option delta: Delta measures the sensitivity of the option's price to changes in the underlying asset price, providing insights into whether early exercise is optimal.
\item Days to ex-dividend date divided by DTE: This ratio indicates how close the ex-dividend date is to the expiration, giving a sense of whether the dividend payout justifies early exercise.
\item Delta multiplied by dividend yield: This interaction term captures the combined effect of price sensitivity and dividend incentives on early exercise decisions.
\end{itemize}

\subsubsection{Liquidity and Market Conditions Module}
The Liquidity and Market Conditions Module is designed to capture how liquidity and broader market conditions impact option pricing and the potential for early exercise. The features used in this module include:
\begin{itemize}
\item Option volume: The number of contracts traded, indicating market interest and liquidity for the option.
\item Bid-ask spread: The difference between the bid (buy) and ask (sell) prices, reflecting the liquidity and transaction costs associated with trading the option.
\item Put-call ratio: The ratio of put options to call options traded, often used as a sentiment indicator of market direction.
\item Log of put-call ratio: A logarithmic transformation to stabilize variance and capture nonlinear relationships between sentiment and option price.
\item VIX: The volatility index, representing market expectations of future volatility and often used as a fear gauge.
\item Volume divided by bid-ask spread: This ratio captures the liquidity of the option, where higher volume relative to spread indicates better liquidity and lower transaction costs.
\item VIX multiplied by implied volatility: This interaction term reflects how market-wide volatility expectations influence individual option pricing.
\item Put-call ratio divided by VIX: This feature captures how sentiment, as reflected by the put-call ratio, interacts with broader market volatility expectations.
\end{itemize}

\subsubsection{Macro-Economic and Sentiment Module}
The Macro-Economic and Sentiment Module is designed to capture how macroeconomic and sentiment-driven features could help predict the likelihood of early exercise and option price movements through broader economic conditions. The features used in this module include:
\begin{itemize}
\item GDP growth: The rate of economic growth, which can affect investor expectations and market performance.
\item Inflation rate: Rising inflation can erode the value of fixed cash flows, impacting options tied to long-term assets.
\item Unemployment rate: A key indicator of economic health, which can influence market sentiment and stock prices.
\item Put-call ratio: Reflects market sentiment regarding the balance of bullish (call) vs bearish (put) expectations.
\item Log of put-call ratio: A transformation to better capture nonlinear relationships in market sentiment.
\item VIX: The volatility index, representing market uncertainty.
\item GDP growth multiplied by inflation rate: This interaction term reflects how economic growth and inflation jointly affect asset prices and option valuations.
\item Unemployment rate divided by GDP growth: This ratio provides insight into the broader economic environment's impact on market sentiment and option prices.
\item VIX multiplied by put-call ratio: This feature captures the relationship between market volatility and sentiment.
\end{itemize}

\subsubsection{Option Greeks and Interaction Module}
The Option Greeks and Interaction Module is designed to capture the intricate relationships between the option Greeks and other factors, ensuring the model can account for the complex sensitivities of American options, especially under varying market conditions. The features used in this module include:
\begin{itemize}
\item Option delta: Measures the sensitivity of the option’s price to changes in the underlying asset price.
\item Option gamma: Captures the rate of change in delta as the underlying asset price changes.
\item Option vega: Measures the sensitivity of the option’s price to changes in implied volatility.
\item Option theta: Represents the rate of time decay in the option’s price.
\item Option rho: Measures the sensitivity of the option’s price to changes in interest rates.
\item Delta multiplied by vega: This interaction term captures how the option’s sensitivity to asset price changes is affected by volatility.
\item Gamma multiplied by theta: This captures how the second-order sensitivity to asset price changes interacts with time decay.
\item Rho multiplied by interest rate: Reflects how the sensitivity to interest rates interacts with the prevailing rate environment.
\end{itemize}

%--- Section ---%
\section{Experiments and Results}
\label{sec:results}

\subsection{Data Pre-Processing}
In order to create datasets necessary for the proposed MNN model, we conducted thorough data pre-processing, ensuring the inclusion of relevant features for each of the six modules. Data was originally collected for AAPL, NVDA and QQQ call options for the year 2023, which included relevant features like quote date, expiration date, underlying price, days to expiration (DTE), option Greeks (delta, gamma, vega, theta, rho), implied volatility, volume, and strike price. Additional fields like the strike distance and moneyness were then calculated through existing features. 

The dataset was further enriched with dividend data sourced from Yahoo Finance. Key macroeconomic indicators like GDP growth, inflation, and unemployment rates were also sourced from the Federal Reserve Economic Data (FRED). Furthermore, interest rates were derived from FRED by matching U.S. Treasury maturities to the option’s DTE. This served as a proxy for the risk-free rate, enhancing the accuracy of the early exercise and dividend-related calculations. Other parameters such as the put-call ratio, intrinsic value, and historical volatility was computed using historical price data. The final pre-processed dataset was tailored to include the essential features for each module, along with newly engineered features such as Gamma divided by DTE, among others to better capture the complexities of option pricing. 

\subsection{Experiments}
To evaluate the effectiveness of our proposed Modular Neural Network (MNN) model in predicting American call option prices, we conducted several experiments. These experiments aimed at evaluating the performance of the MNN model against traditional pricing models and a basic feed-forward neural network (FNN) as a benchmark. The goal was to assess the accuracy of each approach and understand the benefits of using modularization in the context of options pricing.

\subsubsection{MNN Model with Hyper-parameter Tuning}
The initial experiment focused on building the MNN model based on our proposed architecture. First, the data was split into a training and testing dataset and feature scaling was applied using MinMaxScaler. The dataset was further separated into six specialized feature sets corresponding to the different modules: Intrinsic Value and Moneyness, Time Value and Volatility, Early Exercise and Dividend, Liquidity and Market Conditions, Macro-Economic and Sentiment, and Option Greeks and Interaction. The modular approach helped each module learn specific aspects of option pricing independently.

To optimize the MNN architecture, we applied hyper-parameter tuning using Keras Tuner. The hyper-parameter tuning process aimed to find the optimal number of layers, neurons per layer, and activation functions to utilize for each module respectively as well as the final module connected to the output of all six modules. We only conducted hyper-parameter tuning on the AAPL dataset to find the best model, as performing extensive tuning on each dataset would have been computationally expensive and time-consuming. This best model was then used for both NVDA and QQQ datasets. For each of the modules, hyper-parameter tuning was conducted by testing various configurations of one or two hidden layers, with 32, 64, or 128 neurons. Additionally the performance of four activation functions was also investigated: ReLU (Rectified Linear Unit), Exponential Linear Unit (ELU), Tanh (Hyperbolic Tangent), and Swish. For the final model, tuning was expanded to explore combinations of one to three hidden layers with the same set of activation functions. The main criteria for model selection is the minimization of validation loss.

The best MNN model on the AAPL dataset as indicated by the lowest validation error is as follows: 
\begin{itemize}
\item Intrinsic Value and Moneyness Module: 2 layers with 128 neurons and activation functions of Swish for both 
\item Time Value and Volatility Module: 2 layers with 32 and 64 neurons respectively and activation functions of ReLU for both 
\item Early Exercise and Dividend Module: 1 layer with 64 neurons and an activation function of ReLU
\item Liquidity and Market Conditions Module: 2 layers with 128, 64 neurons respectively and activation functions of ReLU for both
\item Macro-Economic and Sentiment Module: 1 layer with 64 neurons and an activation function of Swish
\item Option Greeks and Interaction Module: 2 layers with 128, 32 neurons respectively and activation functions of ELU and Tanh respectively
\item Final Model: 2 layers with 128, 32 neurons respectively and activation functions of ReLU and Tanh respectively
\end{itemize}

\subsubsection{Traditional Pricing Models}
To evaluate the MNN model, we conducted experiments using two traditional American options pricing models: the Barone-Adesi and Whaley (B-AW) Approximation and the Binomial Option Pricing (BOP) Model. 

We computed the B-AW option prices using key features such as the underlying asset price, strike price, interest rate, volatility, and dividend yield. Similarly, the BOP model, implemented with 100 steps, approximated option prices using a discrete-time model that also considered early exercise behavior. Both traditional models provided us with a theoretical option price based on established financial methods.

\subsubsection{FNN Model with Hyper-parameter Tuning}
To further explore the value of our MNN’s modular approach, we constructed a simpler Feed-forward Neural Network (FNN) using a selected subset of important features. These features included fundamental factors such as the underlying asset price, strike price, days to expiration, implied volatility, and option Greeks. Hyper-parameter tuning was also done on this FNN model to find the optimal number of layers, neurons per layer and activation functions used. Similar to the MNN model, hyper-parameter tuning was only done on the AAPL dataset to find the best model, which was then used for both the NVDA and QQQ datasets. Various configurations of one to three hidden layers with 32, 64 and 128 neurons were tested. Additionally, similar to the MNN model, the performance of four activation functions was also investigated: ReLU (Rectified Linear Unit), Exponential Linear Unit (ELU), Tanh (Hyperbolic Tangent), and Swish.

The best FNN model of the AAPL dataset as indicated by the lowest validation error had 3 layers of 64, 128 and 128 neurons respectively and activation functions of ReLU for the first two layers and Tanh for the last layer. 

The FNN experiment provided a useful comparison point by evaluating a traditional neural network approach that lacked the modular specialization found in the MNN. This allowed us to assess the benefits of modularization by comparing how well the FNN captured interactions between features compared to the MNN.

\subsection{Results}
The results of the experiments for predicting American call option prices were evaluated using two key metrics: Root Mean Squared Error (RMSE) and normalized RMSE (nRMSE) by mean. These metrics were compared across the four different models: the Modular Neural Network (MNN), traditional models (Barone-Adesi Whaley (B-AW) and Binomial Option Pricing (BOP)), and a Feed-forward Neural Network (FNN).

\begin{table}[ht]
\centering
\begin{tabular}{||c c c c c||} 
 \hline
  & MNN & B-AW & BOP & FNN \\ [0.5ex] 
 \hline\hline
 RSME & 3.335 & 8.632 & 8.639 & 3.551\\ 
 \hline
 nRMSE & 0.0849 & 0.2191 & 0.2192 & 0.0904\\
 \hline
\end{tabular}
\caption{RMSE and nRMSE for AAPL Stock}
 \label{tab-I}
\end{table}

For AAPL options, the MNN model significantly outperformed both traditional methods and the FNN model as depicted in Table \ref{tab-I}. The MNN achieved an RMSE of 3.335 and an nRMSE of 0.0849, demonstrating its strong predictive performance. In comparison, the traditional methods produced much higher errors, with the B-AW method yielding an RMSE of 8.632 and an nRMSE of 0.2191, while the BOP method had similar values. The FNN model also performed worse than the MNN, with an RMSE of 3.551 and an nRMSE of 0.0904. These results indicate that the MNN model captures the complexities of option pricing for AAPL stock more effectively than the other approaches.

\begin{table}[ht]
\centering
\begin{tabular}{||c c c c c||} 
 \hline
  & MNN & B-AW & BOP & FNN \\ [0.5ex] 
 \hline\hline
 RSME & 17.516 & 41.737 & 41.748 & 19.563\\ 
 \hline
 nRMSE & 0.1701 & 0.4060 & 0.4061 & 0.1900\\
 \hline
\end{tabular}
\caption{RMSE and nRMSE for NVDA Stock}
 \label{tab-II}
\end{table}

For NVDA options, the MNN model again showed superior performance as depicted in Table \ref{tab-II}. The MNN had an RMSE of 17.516 and an nRMSE of 0.1701, which was significantly lower than the errors of the traditional methods. The B-AW method resulted in an RMSE of 41.737 and an nRMSE of 0.4060, while the BOP method produced almost identical results. The FNN model, although better than the traditional methods, still had higher errors than the MNN, with an RMSE of 19.563 and an nRMSE of 0.1900. This confirms the MNN's robustness in predicting NVDA option prices.

\begin{table}[ht]
\centering
\begin{tabular}{||c c c c c||} 
 \hline
  & MNN & B-AW & BOP & FNN \\ [0.5ex] 
 \hline\hline
 RSME & 8.052 & 14.881 & 14.889 & 8.699\\ 
 \hline
 nRMSE & 0.1878 & 0.3489 & 0.3491 & 0.2029\\
 \hline
\end{tabular}
\caption{RMSE and nRMSE for QQQ Stock}
 \label{tab-III}
\end{table}

For QQQ options, the MNN model achieved an RMSE of 8.052 and an nRMSE of 0.1878, outperforming both the traditional methods and the FNN model as depicted in Table \ref{tab-III}. The B-AW method produced an RMSE of 14.881 and an nRMSE of 0.3489, and the BOP method delivered similar results. The FNN model had an RMSE of 8.699 and an nRMSE of 0.2029, indicating that while it performed better than the traditional methods, it still lagged behind the MNN in accuracy.

%--- Section ---%
\section{Discussion}
\label{sec:discussion}
The findings indicate that the Modular Neural Network (MNN) model significantly outperforms both traditional methods (Barone-Adesi Whaley (B-AW) and Binomial Option Pricing (BOP)) and the Feed-forward Neural Network (FNN) in predicting American call option prices. This is further illustrated by the lower RMSE and nRMSE values of the MNN that highlight its ability to capture the complexities of American call option pricing.

The MNN's superior performance across all three stocks— AAPL, NVDA, and QQQ —demonstrates the effectiveness of its modular structure. By breaking down the pricing process into specialized components, the MNN was able to capture non-linear interactions between these variables more effectively than the traditional models as well as the FNN model. The traditional models are based on certain fixed assumptions such as constant volatility which do not account for real-world market dynamics leading to lower accuracies. On the other hand, the MNN is both flexible and specialized which allows it to predict call options more accurately. 

The lower nRMSE values for AAPL stocks in comparison to NVDA and QQQ could be attributed to the fact that both the MNN and FNN model's hyper-parameter tuning was done only on the AAPL dataset. Stock-specific results further emphasize the benefits of modularization. Looking at the results of the NVDA dataset, we can observe that the MNN model significantly outperformed traditional models, with an RMSE of 17.516 compared to over 41 for B-AW and BOP. This suggests that the MNN is more suitable to handle stocks with high volatility and market complexity like NVDA \parencite{chen_comparison_2022}. Even in the case of QQQ, which is highly influenced by market sentiment \parencite{yang_twitter_2015}, the MNN's RMSE of 8.052 was much lower than the RMSE values above 14 for traditional models, and slightly better than the FNN model.

In summary, the MNN’s modular approach proves highly effective in improving predictive performance for American call option prices, providing a clear advantage over traditional models and FNNs. The results confirm the value of modularization in capturing the intricate relationships that drive option pricing.

%--- Section ---%
\section{Conclusion}
\label{sec:conclusion}
The objective of this study was to investigate the accuracy of American call option pricing through a specific type of neural network known as Modular Neural Networks (MNN) and further assess its effectiveness with respect to traditional models as well as Feed-forward Neural Networks (FNN). Through our experiments, it became evident that the MNN model significantly outperformed both the traditional Barone-Adesi Whaley (B-AW) and Binomial Option Pricing (BOP) methods, as well as the simpler FNN model, across all three stocks—AAPL, NVDA, and QQQ.

The modular architecture of the MNN, designed to capture distinct aspects of options pricing, proved to be highly effective. By isolating these factors into specialized modules, the MNN demonstrated an ability to capture the complex, non-linear interactions that drive American options pricing. This structure enabled the MNN to consistently achieve lower RMSE and nRMSE values, confirming its superior predictive power. In comparison, the traditional models struggled with the dynamic market conditions inherent in American options pricing. The FNN, though more adaptable than traditional methods, still could not match the MNN in accuracy, as it lacked the specialized focus of the modular approach.

More recently, advances in explainable AI (XAI) have addressed the "black-box" nature of neural networks, which has been a concern in finance. Transparency is crucial, especially when using machine learning for pricing and risk management. \cite{dixon_machine_2020} introduced interpretability techniques that make neural network models more understandable in financial contexts, particularly in how they handle option pricing. These advancements provide a bridge between the computational power of neural networks and the transparency required in finance.

In conclusion, the MNN architecture offers a powerful and flexible solution for predicting American call option prices, providing clear advantages in accuracy and adaptability over traditional models and simpler neural networks. This study highlights the potential of neural networks, and particularly MNNs, in transforming financial modeling and enhancing predictive accuracy in complex markets like options trading.

%-------------------------------------------
% References
%-------------------------------------------
\break
% Print bibliography
\printbibliography

\end{document}